\documentstyle[12pt]{article}
\begin{document}
\thispagestyle{empty}
\newcommand{\vsl}{v\!\!\!/}
\begin{flushright}
{\sc BU-HEP} 93-28\\December 1993
\end{flushright}
\vspace{.1cm}
\begin{center}
{\LARGE\sc Chiral Perturbation Theory on the Lattice;
Strong Coupling Expansion\footnote{This work was supported in part
under DOE grant DE--FG02--91ER40676, DOE contract DE--AC02--89ER40509,
NSF contract PHY--9057173, and by funds from the Texas National Research
Laboratory Commission under grant RGFY91B6.}}\\[2cm]
{\sc Stanley Myint and Claudio Rebbi}\\[1.5cm]
{\sc Physics Department\\[3mm]
Boston University\\[3mm]
590 Commonwealth Avenue\\[3mm]
Boston, MA 02215, USA}\\[1.5cm]
{\em submitted to Nuclear Physics B}\\[1.5cm]
{\sc Abstract}
\end{center}
\begin{quote}
We evaluate the coefficients of the effective chiral Lagrangian to $O(p^4)$
in the strong coupling, large-N expansion. In this limit we explicitly
perform the functional integral over fundamental degrees of freedom and
obtain the effective chiral Lagrangian. We perform this
calculation on the Body Centered Hypercubical lattice which preserves
Euclidean invariance to order $p^4$. We further discuss how the
coefficients could be obtained numerically, out of the strong-coupling domain.
\end{quote}
\vfill
\newpage

\section{Introduction}

\hspace{6mm}Effective chiral Lagrangians, originally introduced by Weinberg
\cite{wein}, provide a very powerful method to describe the low energy
phenomenology of QCD, exploiting the consequences of chiral symmetry and
current algebra \cite{gellmann}.  The idea
underlying effective Lagrangians is simple and elegant.  There are many
phenomena which, although the fundamental theory involves several
different scales of length and energy, only lead to the excitation of a
subset of all the degrees of freedom.  We will call such degrees of
freedom the effective degrees of freedom.  Under these circumstances it
is conceptually possible to integrate out of the theory all of the
other degrees of freedom and describe the phenomena in terms of an
alternative Lagrangian, namely the effective Lagrangian, more amenable
to calculations.  Thus in low energy QCD, below the threshold for
production of resonances, the effective degrees of freedom are those of
the pseudoscalar fields, i.e. the would be massless Goldstone bosons of
spontaneously broken chiral symmetry, and pion and kaon interactions can
be described in terms of an effective chiral Lagrangian.  Another
example is found in the physics of heavy quarks, where effective
non-relativistic Lagrangians can be used to describe a large class of
processes which do not lead to the excitation of the relativistic degrees of
freedom of the heavy quarks.

        The predictive power of effective Lagrangians in general derives
from the existence of some underlying exact or approximate symmetry,
such as chiral symmetry in QCD, which restricts substantially the type
of couplings that can appear in the Lagrangian itself.  Moreover, these
couplings can be arranged in an expansion into powers of some parameter
related to the suppression of the non-effective degrees of freedom (e.g.
the magnitude of the external momenta in QCD) and at any order in the
expansion there is only a finite number of possible coupling terms.  In
typical phenomenological applications, then, the corresponding coupling
constants are fixed by comparing the results that follow from the
effective theory to a subset of the experimental data.  This determines
the effective Lagrangian, which can subsequently be used to calculate a
variety of other experimental observables \cite{gl1,gl2}.

        Although this way of proceeding provides a very powerful
computational tool, capable of producing a large number of theoretical
predictions starting from general symmetry considerations and a limited
amount of experimental information, it would be even more valuable to be
able to calculate the coefficients of the effective Lagrangian directly
from first principles.  In the case of low energy QCD, this goal appears
within reach of advanced lattice numerical simulations and in this
article we wish to present a framework by which it can be achieved.

        In some sense, once one has the possibility of calculating
observables by means of numerical simulation methods, deriving the
coefficients of the effective Lagrangian might appear straightforward:
one would just need to compare a sufficient number of predictions from
the effective Lagrangian to the results of numerical calculations, very
much as one does comparing them to experimental data in the
phenomenological derivation.  But then, from this point of view, why
resort to an effective Lagrangian at all, if the observables can be
calculated directly?  The point is that numerical simulations provide
only an approximate computational tool, where the accuracy of the
calculations is severely limited by statistical errors and the
discretization procedure.  The precision by which one can determine
quantities such as the coefficients of an effective chiral Lagrangian,
or even the feasibility of such a calculation in absolute, depends
dramatically on the choice of a sound computational technique.

        The basic idea underlying the method we wish to propose is that
the separation into effective and non-effective degrees of freedom of
the continuum theory will be mirrored by a similar distinction in the
theory discretized over a lattice.  It will then be possible to define a
lattice effective Lagrangian to be used as an intermediate step in the
derivation of the coefficients of the continuum chiral Lagrangian from
the fundamental theory discretized over the lattice.  The advantage of
this two step procedure is that the lattice effective Lagrangian will
contain explicitly the collective fields responsible for the long
distance behavior of the fundamental lattice theory.  The calculation of
its coefficients, which for realistic applications will have to be made
through a numerical integration of the non-effective degrees of freedom,
will then only involve shorter scales of length, and should be feasible
on a lattice of moderate size.  Being able to reduce the size of the
lattice, and therefore the total number of variables in a numerical
simulation, produces major advantages for the accuracy and efficiency of
the calculation.

        In this article we will illustrate these ideas by deriving the
lattice effective chiral Lagrangian in the strong coupling approximation
and using it to calculate, always for strong coupling, the coefficients
of the $p^4$ terms of the continuum Lagrangian.  This will be done in
Section 3, after a quick recapitulation, presented in Section 2, of the
form of the continuum chiral Lagrangian and of some features of the
lattice discretization of the QCD fundamental Lagrangian that will be
relevant for the rest of the paper.  Section 3 owes a lot to the truly
pioneering works of Kluberg-Stern, Morel, Napoly and Peterson
\cite{klus} and of Kawamoto and Smit \cite{kaws} where
the methods for integrating over quark and gauge degrees of freedom in
the strong coupling limit were clearly laid out and an effective strong
coupling Lagrangian, expressed in terms of meson fields, was first
derived.  We add to their work, however, the consideration of a
different lattice, namely the Body Centered Hypercubical (BCH) or F4
lattice, whose greater symmetry implies that invariance under lattice
transformations carries over to Euclidean invariance up to the terms of order
$p^4$ in the continuum. This is clearly crucial for a strong
coupling derivation of the $p^4$ terms in the effective chiral
Lagrangian. We were inspired to use this lattice by a talk
presented by Boghosian at the Amsterdam Symposium on Lattice Gauge
Theories. This talk \cite{boghosian} focused on the application of the
BCH lattice in the theory of cellular automata and lattice gases,
however the $p^4$ invariance had been previously emphasized
and exploited by Neuberger \cite{neuberger} in a study of the Higgs
phenomenon. In the context of
lattice field theories, this lattice was also considered in references
\cite{celmaster} -\cite{bhanot}.

        Finally, in Section 4, we will discuss how one could
derive the parameters of the chiral lattice Lagrangian for
more realistic values of the QCD coupling constant, closer to the
continuum limit, by numerical simulations coupled to real space
renormalization techniques.

\section{Chiral Perturbation Theory and Lattice QCD}

\hspace{6mm}For the convenience of the reader and in order to establish
a point of reference we review the basic features of chiral perturbation
theory\footnote{Several reviews are given in
references \cite{georgi} - \cite{meissner}. } and of the lattice
formulation of QCD.

Following Gasser and Leutwyler \cite{gl1} and \cite{gl2},
we introduce the generating functional for the connected Green's functions
of QCD in the presence of scalar, pseudoscalar, vector and axial vector
sources\footnote{Although it is in principle possible to include tensor
sources, they are usually neglected. }
\begin{equation}
e^{i W[s,p,v,a]}=<0\mid T e^{i \int d^4 x \cal L}\mid 0>,
\end{equation}
\begin{equation}
{\cal L}={\cal L}^0_{QCD} -\bar{q}(x)[s(x)-i\gamma_5 p(x)]q(x)
+\bar{q}(x)\gamma^{\mu}[v_{\mu}(x)+\gamma_5 a_{\mu}(x)]q(x). \label{2.2}
\end{equation}
In this expression, $q,\bar{q}$ represent the lightest quark flavor triplet and
${\cal L}^0_{QCD}$ is what remains of the Lagrangian of QCD when the masses
of three lightest quarks are set to zero
\begin{equation}
{\cal L}^0_{QCD}=-\frac{1}{4 g^2} G_{\mu\nu}(x)G^{\mu\nu}(x) +
\bar{q}(x)\gamma^{\mu}[i\partial_{\mu}+ G_{\mu}(x)]q(x).
\end{equation}
$s(x),p(x),v_{\mu}(x)$ and $a_{\mu}(x)$ are Hermitian matrices in flavor
space. The quark mass matrix
\begin{equation}
{\cal M} =
\left( \begin{array}{ccc}
       m_u            \\
             & m_d    \\
             &      & m_s
\end{array}
\right)
\end{equation}
is contained in the scalar field $s(x)$. In this paper we will not
consider the $SU(3)$ singlet vector and axial vector current:
\begin{equation}
{\rm Tr}[v_{\mu}(x)]={\rm Tr}[a_{\mu}(x)]=0.
\end{equation}
It is convenient to define the right and left handed currents:
\begin{equation}
r_{\mu}(x)\equiv v_{\mu}(x)+a_{\mu}(x),\quad
l_{\mu}(x)\equiv v_{\mu}(x)-a_{\mu}(x).
\end{equation}
The Lagrangian (\ref{2.2}) exhibits a local flavor chiral
symmetry $SU_L(3)\times SU_R(3)$:
\begin{eqnarray}
q(x) & \rightarrow &
L(x)\frac{1-\gamma_5}{2}q(x)+R(x)\frac{1+\gamma_5}{2}q(x),
\nonumber \\
r_{\mu}(x) & \rightarrow &
R(x) r_{\mu}(x) R(x)^{\dagger} +i
R(x)\partial_{\mu}R(x)^{\dagger},
\nonumber \\
l_{\mu}(x) & \rightarrow &
L(x) l_{\mu}(x) L(x)^{\dagger} +i
L(x)\partial_{\mu}L(x)^{\dagger},
\nonumber \\
s(x)+i p(x) & \rightarrow & R(x)[s(x)+i p(x)]L(x)^{\dagger}, \label{2.6}
\end{eqnarray}
with matrices $L(x),R(x) \in SU(3)$.

The goal of chiral perturbation theory is to produce an expansion of
the generating functional $W$ in powers of the external
momenta and quark masses.  Because of the presence of pseudo
Goldstone bosons one can not simply perform
a Taylor series expansion of $W$, but one must introduce suitable fields
to account for their degrees of freedom.  These fields
are collected in a unitary $3\times3$ matrix:
\begin{equation}
U(x)=\exp{[\frac{i\pi^a(x)\lambda^a}{F^0}]}, \quad U(x)U(x)^{\dagger}=1.
\end{equation}
$\lambda^a,a=1,...,8 $ are Hermitian generators of $SU(3)$, with the
normalization
\begin{equation}
{\rm Tr}(\lambda^a \lambda^b)=2\delta^{ab},
\end{equation}
$\pi^a(x)$ are real pseudoscalar fields and $F^0$ is the pion decay
constant in the chiral limit:
\begin{eqnarray}
<0 \mid \bar{q}(x)\gamma_{\mu}\gamma_5 \frac{\lambda^a}{2}q(x)\mid \pi^b>
& = & i f_{\pi} p_{\mu} \delta^{ab}, \nonumber \\
f_{\pi}=F_0(1+O(m_q))& \approx & 93.3 MeV .
\end{eqnarray}
To the lowest order $O(p^2)$, one must only evaluate tree level
diagrams with
a classical Lagrangian involving terms with two derivatives
and one power of scalar and pseudoscalar sources:
\begin{equation}
{\cal L}_2=\frac{F_0^2}{4}{\rm Tr}\{[D_{\mu}U(x)] [
D^{\mu}U(x)]^{\dagger}+\chi(x) U(x)^{\dagger}+\chi(x)^{\dagger} U(x)\}.
\label{2.10}
\end{equation}
In this expression,
\begin{equation}
D_{\mu}U(x)=\partial_{\mu}U(x)-i
r_{\mu}(x) U(x) +i U(x) l_{\mu}(x) \label{2.11}
\end{equation}
is a flavor covariant derivative and matrix $\chi(x)$ collects the scalar
and pseudoscalar fields:
\begin{equation}
\chi(x)=2 B_0[s(x)+i p(x)].
\end{equation}
The constant $B_0$ is related to the quark condensate in the chiral
limit:
\begin{equation}
<0\mid \bar{u} u\mid0>_0=<0\mid \bar{d} d\mid0>_0=<0\mid \bar{s}
s\mid0>_0=-F_0^2 B_0(1+O(m_q)).
\end{equation}
In the expansion in powers of pion momenta, vector and axial vector
sources count as contributions of order $p$, whereas scalar and
pseudoscalar sources
count as contributions of order $p^2$.

To the next to leading order $O(p^4)$, one encounters two kinds of
contributions: \begin{itemize}
\item{one loop diagrams with ${\cal L}_2$ vertices;}
\item{tree diagrams with ${\cal L}_4$ vertices.}
\end{itemize}
${\cal L}_4$ is the most general Lagrangian which contains four powers of
momenta consistent with the local chiral flavor symmetry (\ref{2.6}).
Therefore, it can be formed only from the covariant derivatives (\ref{2.11})
and the field strengths for right and left-handed fields:
\begin{eqnarray}
F^{\mu\nu}_R&=&\partial^{\mu}r^{\nu}-\partial^{\nu}r^{\mu}-
i[r^{\mu},r^{\nu}] , \nonumber \\
F^{\mu\nu}_L&=&\partial^{\mu}l^{\nu}\,-\partial^{\nu}l^{\mu}\:-
i[l^{\mu},l^{\nu}] .
\end{eqnarray}
\begin{eqnarray}
{\cal L}_4 & = & L_1 \{{\rm Tr}[(D_{\mu}U)(D^{\mu}U)^{\dagger}]\}^2 + L_2
{\rm Tr}[(D_{\mu}U)^{\dagger} (D_{\nu}U)]{\rm Tr}[(D^{\mu}U)^{\dagger}
(D^{\nu}U)] \nonumber \\ &  +&
L_3 {\rm Tr}[(D_{\mu}U)^{\dagger}(D^{\mu}U)(D_{\nu}U)^{\dagger}(D^{\nu}U)] +
L_4 {\rm Tr}[(D_{\mu}U)^{\dagger}(D^{\mu}U)] {\rm Tr}(\chi^{\dagger}U + \chi
U^{\dagger}) \nonumber \\ &  +&
L_5 {\rm Tr}[(D_{\mu}U)^{\dagger}(D^{\mu}U)
(\chi^{\dagger}U + \chi U^{\dagger})
]+ L_6[{\rm Tr}(\chi^{\dagger}U + \chi U^{\dagger})]^2
\nonumber \\ &  +& L_7
[{\rm Tr}(\chi^{\dagger}U - \chi U^{\dagger})]^2
+L_8 {\rm Tr}(\chi^{\dagger} U\chi^{\dagger} U+\chi U^{\dagger}\chi
U^{\dagger}) \nonumber \\ &  -&
i L_9 {\rm Tr}[F^R_{\mu\nu}(D^{\mu}U)(D^{\nu}U)^{\dagger} +
F^L_{\mu\nu}(D^{\mu}U)^{\dagger} (D^{\nu}U)]
+L_{10} {\rm Tr}(U^{\dagger}F^R_{\mu \nu}U F^{L \mu \nu})
\nonumber \\ &  +&
H_1 {\rm Tr}(F^R_{\mu \nu}F^{R \mu \nu} + F^L_{\mu \nu}F^{L \mu \nu})
+H_2 {\rm Tr}(\chi^{\dagger} \chi).
\end{eqnarray}

As we see, at leading order, chiral symmetry restricts the number of terms in
the effective chiral Lagrangian to only two, so that one needs to know only two
constants, $F_0$ and $B_0$, in order to describe all low energy phenomena
in QCD to order $p^2$. At the next to leading order one needs ten more
constants: $L_1$ - $L_{10}$.
(The terms with coefficients $H_1$ and $H_2$ are of no physical
significance and we will disregard them from now on.)

In two seminal papers by Gasser and Leutwyler \cite{gl1,gl2}
these ten coefficients have been calculated to reasonable accuracy by
comparison with experimental data. It would be of course important
to be able to determine all the coefficients theoretically.  This
requires a non-perturbative calculation and lattice numerical
simulations represent today the only method by which such a
calculation could be carried out from first principles.  Otherwise
one can try to calculate the coefficients from models which are
similar to, but easier to solve than QCD.
For an account of various attempts at a theoretical determination
of the coefficients see, for example,
\cite{holdom,ebert,ecker} and references
therein. In recent years, a rather exhaustive calculation has appeared in
references \cite{espriu,bruno}.

Our goal is to provide a framework by which the coefficients
of the terms of order $p^4$ in the chiral Lagrangian can be
efficiently calculated by means of a lattice numerical simulation.
The idea is to proceed through the intermediary of an effective
chiral Lagrangian defined over the lattice.  The problem then
splits into two separate calculations:
\begin{itemize}

\item{A numerical simulation is used to calculate the coefficients
of the lattice chiral Lagrangian through a matching of lattice
observables.  As we will argue in Sect.~4, it should be possible
to carry out this step on a reasonably small lattice and, therefore,
with high statistical accuracy.}

\item{The coefficients of the continuum chiral Lagrangian will be
obtained from those of the lattice effective Lagrangian by
an expansion of the observables for small lattice momenta and
by carrying out the appropriate renormalizations.  This step
can be performed analytically, by a perturbative expansion
of the lattice chiral Lagrangian, if the coefficients, as
determined in the previous step, are such as to validate
the expansion.  Otherwise it would have have to be done numerically.}

\end{itemize}

We believe that one can carry out these two steps with better accuracy
than a direct determination of the coefficients of the
continuum Lagrangian through a comparison with observables
calculated over the lattice.

While in Sect.~4 we plan to discuss in some detail the procedures
that could be followed to perform such calculations in the physically
interesting domain where continuum
scaling sets in, the main result which we wish to present in this
paper is an illustration of the method in the strong coupling
approximation.  Precisely, we wish to establish the possibility of
defining a lattice chiral Lagrangian by deriving it explicitly
at leading order in the strong coupling expansion in
$\beta = 6/g_0^2$.  An expansion for low
lattice momenta will then allow us to identify the values of the
coefficients of the continuum Lagrangian.

In the balance of this section we review a few basic features of the
lattice regularization of QCD that will be useful for our
subsequent derivation of the lattice chiral Lagrangian and will
also serve to establish our notation.

In the lattice regularization of QCD, Euclidean space-time is
generally discretized
by the introduction of a regular lattice of points
and oriented links joining neighboring points.  In the vast
majority of applications this lattice is a hypercubical lattice,
however, for reasons that will become clear in Sect.~3, we will
need to work with a different lattice to carry out the strong
coupling expansion.  Therefore we will leave the detailed geometry
of the lattice unspecified for now, but we will assume that
the lattice is regular and uniform, with lattice spacing
$a$.

We will denote by $x$ the coordinates of the lattice points
and by $v$ the fundamental displacement vectors in the lattice.
Thus, for clarification, with a hypercubical lattice we would have
$x=(i_1 a, i_2 a,i_3 a,i_4 a)$ with integer $i_1, i_2,i_3,i_4 $
and $v$ would range over $ (a,0,0,0),(0,a,0,0),(0,0,a,0),(0,0,0,a)$.

The fundamental variables are the gluon fields and the quark
fields.  The gluon fields are finite elements of color $SU(3)$,
which are associated to the oriented links of the lattice.
It is rather standard to denote such variables by $U_{x,v}$,
but eventually, following another standard notation, we will
want to reserve the $U$ symbol for the chiral fields.  We will
therefore denote the lattice gauge fields by $G_{x,v}$.

The gauge part of the action is constructed from the plaquette
variables, i.e. the transport factors around the elementary
closed contours of the lattice.  The plaquettes are defined
by two displacement vectors, and so we will use the notation
$G_{x,v_1 v_2}$ to denote a plaquette variable.  (For the
hypercubical lattice $G_{x,v_1 v_2}=G_{x,v_1}G_{x+v_1,v_2}
G_{x+v_2,v_1}^{\dagger}G_{x,v_2}^{\dagger}$). The gauge part of the
action is then given by
\begin{equation}
S_g= \beta K_1\sum_{x,v_1,v_2}
{\rm Tr}\,(I- G_{x,v_1 v_2}), \label{3.2}
\end{equation}
where we are summing over all plaquettes with both orientations,
$K_1$, as well as $K_2$ in the equation below,
are lattice dependent normalization factors,
\footnote{$K_1=K_2=1/2$ for the hypercubical lattice.}
and the coupling parameter $\beta$ is
related to the bare coupling constant by $\beta=6/g_0^2$.

The quark and antiquark fields,\footnote{Color,
spin and flavor indices will be left implicit where it doesn't
affect the clarity of the formula.} $\psi_x, \bar{\psi}_x$ are
defined over the sites of the lattice. The interaction term between
quarks and gluons is given by:
\begin{equation}
S_{\psi}= a^3 K_2 \sum_x \sum_v (\bar{\psi}_x \vsl
G_{x,v}\psi_{x+v}-\bar{\psi}_{x+v}\vsl G_{x,v}^{\dagger}\psi_x)
\label{3.3}
\end{equation}

We will also include scalar and pseudo-scalar source terms through
a matrix $j_x$:
\begin{equation}
S_j= a^4 \sum_x  \bar{\psi}_x j_x \psi_x
= a^4 \sum_x  \bar{\psi}_x (s_x-i\gamma_5 p_x)\psi_x.\label{3}
\end{equation}

To summarize, equations (\ref{3.2}-\ref{3}) define the QCD action on any
regular uniform lattice.

\section{Strong coupling expansion on the BCH Lattice}

\hspace{6mm}In order to derive a lattice chiral Lagrangian one must integrate
out the high energy degrees of freedom, i.e. the quantum fluctuations of
the gauge and quark fields, re-expressing the generating functional
in terms of pseudoscalar fields conjugate to the external sources.
Thus one effectively performs a bosonization of the original theory.
For general values of the bare lattice coupling constant this can
only be done numerically, but in the strong coupling and in the
large $N$ (number of colors) limits the bosonization can be done analytically.
The techniques for deriving an effective Lagrangian in the strong coupling
limit were laid down in papers by Kluberg-Stern, Morel,
Napoly and Petersson \cite{klus} and Kawamoto and Smit \cite{kaws}.
We add, however, to
the work of these authors an ingredient that plays a crucial role
for the possibility of obtaining the terms of order $p^4$ in the chiral
Lagrangian. As in most lattice calculations, \cite{klus,kaws} deal with
a hypercubical (HC) lattice.  While such a lattice obviously does not have
the symmetry of the continuum, all tensors of rank
up to two symmetric under the group of lattice transformations are also
symmetric under the full group of 4-dimensional rotations of Euclidean
space-time.  However, this property does not carry through to tensors of
higher order, so that, although one could expand the chiral
Lagrangian derived on
a HC lattice to terms of $4^{th}$ order in lattice momentum, one
would obtain terms that cannot be identified with continuum
counterparts. This precludes the possibility of recognizing the
structure of the continuum chiral Lagrangian to $O(p^4)$ on the HC lattice.

In order to pursue our calculation, we will study QCD on a different
lattice, namely the Body Centered Hypercubical (BCH) or F4 lattice,
which, as already mentioned in the introduction,
has a remarkable property that not only tensors of rank two but also
those of rank four have the continuum symmetry.

The BCH lattice can be formed in two equivalent ways. One is to
visualize it as the usual HC lattice from which all the sites with
$\sum_{\mu} x_{\mu} = odd$ (or $even$) have been removed. An equivalent way
is to take all the sites of the HC lattice together with centers of its
elementary cells (hence the name BCH) to represent the sites of the new
lattice. The lattice formed in such a way has the largest symmetry group
of all four dimensional regular lattices. Whereas the usual HC lattice
has 384 element symmetry group, the symmetry group of the BCH lattice has
1152 elements. A consequence of this is the required $p^4$ invariance.

On the BCH lattice every site has 24 nearest neighbors forming
triangular plaquettes. In terms of unit
vectors of the hypercubical lattice $\vec{e}_i$, the links in the
positive direction $v_{ij}^{\alpha}$ are given by
the following twelve linear combinations:
\begin{equation}
v_{ij}^{\alpha} = \frac{\vec{e}_i+\alpha\vec{e}_j}{\sqrt{2}},
\{1\leq i < j \leq 4, \alpha = \pm1\} .\label{17}
\end{equation}

        The use of the BCH lattice entails another advantage in the strong
coupling limit.  It is of rather technical nature and so we will mention
it only briefly.  The definition of lattice fermions encounters some
notorious problems.  A straightforward discretization of the Dirac
operator leads to the introduction of additional poles in the propagators
of the fermions at the corners of the Brillouin zone (species doubling).
Although these extra fermionic modes would in general be coupled through
the gluonic quantum fluctuations, on a hypercubical lattice there is a
fourfold degeneracy of totally decoupled
fermionic degrees of freedom.  This is akin to starting with one flavor of
fermions, but discovering that the theory actually contains four decoupled
flavors.  As a consequence, the spontaneous breaking of chiral symmetry
gives origin to 16 pseudo Goldstone bosons, even though one started with a
single flavor.  On a BCH lattice, because of the higher coordination
number of its vertices, such a separation of the fermionic fields into
four decoupled components (which is not demanded by any general theorem
on lattice fermions) does not take place and one maintains the
relationship between the number of explicit flavor indices and the number
of pseudo Goldstone bosons proper of the continuum theory.\footnote{It
has been argued in Ref.~\cite{celkrausz} that the use of the BCH lattice
actually worsens the problem of lattice fermions as one tries to recover
the continuum limit. This is irrelevant for our purposes, because
our motivation for its use comes solely from the need of obtaining
better symmetry properties in the strong coupling limit.  We would not
advocate using the BCH lattice in numerical simulations done in the
scaling regime, where there is good evidence for the recovery of
rotational symmetry with the ordinary HC lattice.}

In order to study interactions with external vector and axial vector
fields, we will introduce
traceless Hermitian sources on the positive link $v$
starting from a lattice point $x$:
\begin{equation}
\tilde{v}_{x,v} = v_{x,v}^a \lambda^a,\quad
\tilde{a}_{x,v} = a_{x,v}^a \lambda^a.
\end{equation}
We couple these sources to quark and gluon fields in a nonlinear
way through an auxiliary field
$Y(\tilde{v},\tilde{a})$:
\begin{equation}
Y(\tilde{v},\tilde{a})_{x,v}
=\exp\{-\frac{i a}{2}[\tilde{v}_{x,v}+\tilde{v}_{x+v,v}+\gamma_5 (
\tilde{a}_{x,v}+\tilde{a}_{x+v,v})]\}.
\end{equation}
In the classical continuum limit we require that:
\begin{equation}
\tilde{v}_{x,v} \rightarrow \frac{\tilde{v}_i+\alpha
\tilde{v}_j}{\sqrt{2}},\quad
\tilde{a}_{x,v} \rightarrow \frac{\tilde{a}_i+\alpha
\tilde{a}_j}{\sqrt{2}}. \label{3.9}
\end{equation}
The interaction term with couplings to external vector and
axial vector sources is given by:
\begin{equation}
S_{\psi}= a^3 K_2 \sum_x \sum_v (\bar{\psi}_x\vsl
G_{x,v} Y_{x,v}\psi_{x+v}-\bar{\psi}_{x+v}\vsl G_{x,v}^{\dagger}
Y_{x,v}^{\dagger}\psi_x). \label{3.10}
\end{equation}

Our final expression for the lattice action is:
\begin{equation}
S=S_g+S_{\psi}+S_j, \label{3.12}
\end{equation}
with $S_g$, $S_{\psi}$ and $S_j$ given in (\ref{3.2}), (\ref{3.10}) and
(\ref{3}).

We set $K_1=K_2=1/12$ and in the limit of vanishing lattice spacing
$a \to 0$ identify the gauge variables with continuum transport
factors through $ G_{x,v}=\exp[i g_0 A_{\mu}(x) v^{\mu}] $. With this
choice, the action (\ref{3.12}) has a correct classical
continuum limit.

It is worthwhile noting that our action possesses chiral flavor gauge
symmetry under transformations:
\begin{eqnarray}
\psi_x & \rightarrow &
(L_x\frac{1-\gamma_5}{2}+R_x\frac{1+\gamma_5}{2})\psi_x,
\nonumber \\
\bar{\psi}_x & \rightarrow &
\bar{\psi}_x (L_x^{\dagger}\frac{1+\gamma_5}{2}+
R_x^{\dagger}\frac{1-\gamma_5}{2}),
\nonumber \\
Y_{x,v}& \rightarrow & (L_x \frac{1-\gamma_5}{2} + R_x
\frac{1+\gamma_5}{2} ) Y_{x,v} (L_{x+v}^{\dagger}\frac{1-\gamma_5}{2} +
R_{x+v}^{\dagger} \frac{1+\gamma_5}{2} ),
\nonumber \\
s_x +i p_x & \rightarrow & R_x (s_x +i p_x)L_x^{\dagger}, \label{30}
\end{eqnarray}
which reduces to the form (\ref{2.6}) in the classical continuum limit.

Generating functional on the lattice can be written as a path integral
over the quark and gluon fields:
\begin{equation}
Z(s,p,v,a)=\int D\psi D\bar{\psi} DG e^{-S(\psi,\bar{\psi},G;s,p,v,a)}.
\end{equation}
In this expression,
$s,p,v,a$ are external sources defined earlier, the action $S$ is given in
(\ref{3.12}) and the integration is performed over all the lattice
variables
(for example: $DG \equiv \prod_{x,v} dG_{x,v}$).

In the strong coupling expansion, the gauge term (\ref{3.2}) is
suppressed by $1/g_0^2$. Therefore, in the leading order we can neglect it,
and the action reduces to
$S=S_{\psi}+S_j$. Now the gauge variables appear only in
$S_{\psi}$, moreover they appear only linearly. This implies that the
product of integrals over gauge variables $DG$
factors into a product of one-link integrals:
\begin{equation}
Z=\int D\psi D\bar{\psi} e^{-S_j} \prod_{x,v} \int dG_{x,v} e^{-W_{x,v}}.
\end{equation}
Here we have defined:
\begin{equation}
W_{x,v} = \frac{a^3}{12} ( \bar{\psi}_x \vsl G_{x,v} Y_{x,v}
\psi_{x+v} - \bar{\psi}_{x+v} \vsl G_{x,v}^{\dagger} Y_{x,v}^{\dagger}
\psi_x).
\end{equation}

One link integrals of this type have been calculated for the case when
the gauge field couples to a bosonic source in \cite{bregross}. In
\cite{klus} it was shown that this result applies equally to an
interaction of gauge with fermionic fields. The interested reader should
consult reference \cite{klus} for the details. It is important to
mention that the bosonization procedure can be
extended to the next order in $1/g_0^2$ expansion \cite{ichinose} and to
arbitrary (odd) $N$ \cite{azakov}. Here we just give the final result
for the generating functional to the leading order in the strong
coupling, large-N limit:
\begin{equation}
Z=\int D\psi D\bar{\psi} e^{N \sum_{x,v} {\rm Tr}[F(\lambda(x,v))]-S_j}.
\label{3.18}
\end{equation}
$\lambda(x,v)$ is a matrix in Dirac-flavor space, defined by its
elements\footnote{Greek letters will denote indices in the direct
product of Dirac and flavor spaces, and Latin letters the color indices.
In this section we will take the number of flavors to be $n$ and only at
the end specialize to $n=3$.}:
\begin{equation}
\lambda(x,v)_{\alpha\beta}=\frac{a^6}{36 N^2} \sum_{i,j=1}^N
\sum_{\delta,\gamma,\epsilon=1}^{4 n}[\bar{\psi}_i^{\alpha}(x)
\psi_i^{\delta}(x) (\vsl Y_{x,v}^{\dagger})_{\gamma\delta}
\bar{\psi}_j^{\gamma}(x+v) \psi_j^{\epsilon}(x+v) (\vsl
Y_{x,v})_{\beta\epsilon}].
\end{equation}
$F(\lambda)$ is the following matrix function\footnote{$F(\lambda)$
should be understood as a power series in Grassmann variables.}
 of $\lambda$:
\begin{equation}
F(\lambda)=1-\sqrt{1-\lambda}+\ln\frac{1+\sqrt{1-\lambda}}{2},
\end{equation}

Next we proceed to convert the fermionic path integral into one over
bosonic variables as in references \cite{klus}, \cite{kaws}. In the
expression (\ref{3.18}), $N {\rm Tr}[F(\lambda)]$ is the interaction term. It
is important to notice that it depends only on color singlet
combinations $\bar{\psi}_i^{\alpha}(x) \psi_i^{\delta}(x)$ at different
points on the lattice. Since these appear also in the source term $S_j$,
we can write:
\begin{equation}
\frac{a^3}{N} \sum_i [\bar{\psi}_i^{\alpha}(x) \psi_i^{\delta}(x)] =
\frac{1}{N a} \frac{\partial}{\partial j_x^{\alpha \delta}}[a^4
\bar{\psi}_x^{\alpha} j_x^{\alpha \delta} \psi_x^{\delta}]=\frac{1}{N
a} \frac{\partial S_j}{\partial j_x^{\alpha \delta}}.
\end{equation}
Therefore we can rewrite the full generating functional as:
\begin{equation}
Z=e^{ N \sum_{x,v} {\rm Tr}[F(\lambda(x,v))] } \int D\psi D\bar{\psi}
e^{-S_j}, \label{3.22}
\end{equation}
with:
\begin{equation}
\lambda(x,v)_{\alpha\beta}=\frac{1}{36 (N a)^2}
\sum_{\delta,\gamma,\epsilon=1}^{4 n}[ \frac{\partial}{\partial
j_x^{\alpha\delta}} (\vsl Y_{x,v}^{\dagger})_{\gamma\delta}
\frac{\partial}{\partial j_{x+v}^{\gamma\epsilon}}(\vsl
Y_{x,v})_{\beta\epsilon} ]. \label{3.22p}
\end{equation}
The path integral over fermionic variables
factorizes into a product of one-site integrals:
\begin{equation}
Z_0\equiv\int D\psi D\bar{\psi} e^{-S_j}=\prod_x \int d\psi_x
d\bar{\psi}_x e^{-a^4 (\bar{\psi}_x j_x \psi_x)}=\prod_x z_0(j_x),
\end{equation}
\begin{equation}
z_0(j_x)\equiv\int d\psi_x d\bar{\psi}_x e^{-a^4 (\bar{\psi}_x j_x \psi_x)}.
\end{equation}
In \cite{klus,kaws} it was
shown that $z_0(j_x)$ can be written as an integral over bosonic variables:
\begin{equation}
z_0(j_x)=\int d{\cal M}_x e^{N {\rm Tr}(a j_x {\cal M}_x)-N
{\rm Tr}(\ln {\cal M}_x) + const}.
\end{equation}
Here ${\cal M}_x$ is a unitary bosonic matrix in Dirac-flavor space, the
trace is over Dirac-flavor indices and the constant is irrelevant so it
will be omitted hereafter.

Using (\ref{3.22}) and (\ref{3.22p}) we obtain an expression for the full
generating functional:
\begin{equation}
Z=\int D{\cal M} e^{ N \{\sum_{x,v} {\rm Tr}[ F(\lambda_{x,v}) ] + \sum_x
{\rm Tr}( a j_x {\cal M}_x) - \sum_x {\rm Tr}(\ln {\cal M}_x) \} }
\equiv \int D {\cal M} e^{-S({\cal M})},\label{3.29}
\end{equation}
where $\lambda$ is now a bosonic matrix:
\begin{equation}
\lambda(x,v)=\frac{1}{36} \vsl Y_{x,v} {\cal M}_{x+v} \vsl
Y_{x,v}^{\dagger} {\cal M}_x.\label{31}
\end{equation}

The transformation of ${\cal M}$ under chiral gauge symmetry follows
from (\ref{30}):
\begin{equation}
{\cal M}_x \rightarrow (L_x \frac{1-\gamma_5}{2} + R_x
\frac{1+\gamma_5}{2} ) {\cal M}_x (L_x^{\dagger}\frac{1+\gamma_5}{2} +
R_x^{\dagger} \frac{1-\gamma_5}{2} ).\label{45}
\end{equation}
One can see that this, together with a law of transformation of
$Y_{x,v}$ (\ref{30}) makes the action $S({\cal M})$ chirally gauge
invariant.
Now we look for a translationally invariant saddle point of the action
$S({\cal M})$ of the form:
\begin{equation}
{\cal M}_x^0=u_0, \label{4.3}
\end{equation}
$u_0$ being proportional to the unit matrix in Dirac-flavor space. From
(\ref{31}), using the fact that $\vsl^2=1$, we obtain
$\lambda_0 \equiv \lambda(x,u_0)=u_0^2/36$.
For a flavor diagonal mass matrix $j_x=s_x=\bar{m}$ and no other
external sources, the action at this
point reduces to:
\begin{equation}
S^0=-N \sum_{x,v}  {\rm Tr}[ F(\frac{u_0^2}{36}) - \frac{1}{12} \ln u_0
+\frac{1}{12} a \bar{m} u_0 ].
\end{equation}
The effective potential is given by:
\begin{equation}
V_{eff}=-\frac{S({\cal M}_x^0)}{volume}.
\end{equation}
By requiring:
\begin{equation}
\frac{dV_{eff}}{d u_0}=0,
\end{equation}
we find the saddle point as a
positive solution of the quadratic equation:
\begin{equation}
u_0=\frac{ -22 (a\bar{m}) \pm \sqrt{ (24 a\bar{m})^2 + 1472} }
{16 + (a\bar{m})^2} \label{4.8}
\end{equation}
For the massless case $u_0=\sqrt{23}/2\approx 2.40$.

We will parameterize the field ${\cal M}_x$ in a nonlinear
way \cite{kaws}:
\begin{equation}
{\cal M}_x=u_0 \exp{ [\frac{i}{F_0}(S_x+\pi_x \gamma_5 + V_x^{\nu}
\gamma_{\nu} + i A_x^{\nu} \gamma_{\nu} \gamma_5 + \frac{1}{2}
T_x^{\nu\rho}\sigma_{\nu\rho})]}.
\end{equation}
Here $S,\pi,V^{\nu},A^{\nu}$ and $T^{\nu\rho}$ are sixteen
Hermitian matrices in flavor space. Of these fields, only $\pi$ are
real Goldstone bosons, as we will show in appendix A.

In principle, we need to integrate out the scalar, vector, axial vector
and tensor fields in order to obtain an effective action for pseudoscalar
mesons. In the first approximation we will neglect these
contributions\footnote{They
are suppressed by the masses of relevant resonances.}, and concentrate
only on the direct interactions among Goldstone bosons:\footnote{In
this paper we will not
consider the effects of the $U(1)$ anomaly in any detail, so we will
restrict our attention to the traceless part of $\pi_x$.}

\begin{equation}
{\cal M}_x=u_0 \exp{(i \frac{\pi_x \gamma_5}{F_0})}.
\end{equation}

Next we expand the action $S({\cal M})$ up to fourth order in Taylor
series around the vacuum (\ref{4.3}) and (\ref{4.8}).
This is sufficient to extract
coefficients of effective chiral Lagrangian to order $p^4$ since the
deviation from the vacuum $\lambda_{x,v}-\lambda_0$ is of $O(p)$.

The action in (\ref{3.29}) can be written as:
\begin{equation}
S = -N \sum_{x,v} {\rm Tr}[F(\lambda_{x,v})] - N u_0 \sum_x
{\rm Tr}(a j_x {\cal M}_x).
\label{4.11}
\end{equation}

We expand ${\rm Tr}[F(\lambda_{x,v})]$ to the fourth order:
\begin{equation}
{\rm Tr}[F(\lambda_{x,v})]=\sum_{n=1}^4\{\frac{1}{n!}\left(\frac{\partial^n
F}{\partial \lambda^n}\right)_{\lambda_0}
{\rm Tr}[(\lambda_{x,v}-\lambda_0)^n]\}.\label{41}
\end{equation}
We evaluate the Dirac traces in a basis where $\gamma_5={\rm
diag}(1,1,-1,-1)$. By using the relations
\begin{eqnarray}
\vsl {\cal M}_{x+v} & = &{\cal M}_{x+v}^{\dagger} \vsl, \nonumber \\
\vsl Y(\tilde{v},\tilde{a})_{x,v}&=&Y(\tilde{v},-\tilde{a})_{x,v}\vsl,
\end{eqnarray}
which follow from definitions of ${\cal M}$ and $Y(\tilde{v},\tilde{a})$,
we can take the trace over spin indices to obtain:
\begin{equation}
{\rm Tr}[(\lambda_{x,v}-\lambda_0)^n]=2 \lambda_0^n
{\rm Tr}_f[(Z_{x,v} U_{x+v}^{\dagger} W_{x,v}^{\dagger} U_x-1)^n +
\;{\rm h.c.}\;].
\end{equation}
Here we have defined new variables which are matrices in flavor space only:
\begin{eqnarray}
U_x&\equiv&\exp{(i \frac{\pi_x}{F_0})}, \nonumber \\
W_{x,v}&\equiv&\exp{\{\frac{-i a}{2}[\tilde{v}_{x,v}+\tilde{v}_{x+v,v}+(
\tilde{a}_{x,v}+\tilde{a}_{x+v,v})]\}}, \nonumber \\
Z_{x,v}&\equiv&\exp{\{\frac{-i a}{2}[\tilde{v}_{x,v}+\tilde{v}_{x+v,v}-(
\tilde{a}_{x,v}+\tilde{a}_{x+v,v})]\}}. \label{4.16}
\end{eqnarray}
${\rm Tr}_f$ indicates a trace over flavor indices, and from now on we
will omit the subscript $f$.
Evaluating the second term in (\ref{4.11}) is straightforward and one
obtains the expression for the action in terms of the new variables:
\begin{eqnarray}
S= &-& 2 N \sum_{x,v}
\sum_{n=1}^4\frac{\lambda_0^n}{n!}\left(\frac{\partial^n F}
{\partial \lambda^n}\right)_{\lambda_0}
{\rm Tr}[(Z_{x,v} U_{x+v}^{\dagger} W_{x,v}^{\dagger} U_x-1)^n +
\;{\rm h.c.}\;] \nonumber \\
&-& 2 N u_0 a \sum_x {\rm Tr}[(s_x-i p_x) U_x +\;{\rm h.c.}\;].
\end{eqnarray}
Relations (\ref{30}), (\ref{45}) and (\ref{4.16}) determine the
effect of chiral flavor symmetry transformations on the new
variables\footnote{Note that the usual transformation law for $U$
fields follows from the initial form of transformation of quark fields
(\ref{30}) and the sequence of changes of variables that we performed.}:
\begin{eqnarray}
U_x &\rightarrow & R_x U_x L_x^{\dagger}, \nonumber \\
W_{x,v} &\rightarrow & R_x W_{x,v} R_{x+v}^{\dagger}, \nonumber \\
Z_{x,v} &\rightarrow & L_x Z_{x,v} L_{x+v}^{\dagger}. \label{4.17}
\end{eqnarray}
The next step is to re-express ${\rm Tr}[(\lambda_{x,v}-\lambda_0)^n]$
in terms of
a more convenient variable which will facilitate expansion in powers of
momentum (or equivalently powers of lattice spacing $a$).
The new variable is $\alpha$, defined as:
\begin{eqnarray}
\alpha &\equiv&(\triangle_v U_x)^{\dagger}(\triangle_v U_x),\nonumber \\
\triangle_v U_x &\equiv& W_{x,v}^{\dagger} U_x - U_{x+v} Z_{x,v}^{\dagger}.
\label{61}
\end{eqnarray}
One can express the traces ${\rm Tr}[(\lambda_{x,v}-\lambda_0)^n]$ in terms of
$\alpha$:
\begin{eqnarray}
{\rm Tr}[\lambda_{x,v}-\lambda_0]&=&-2 \lambda_0 {\rm Tr}(\alpha),\nonumber \\
{\rm Tr}[(\lambda_{x,v}-\lambda_0)^2]&=&2 \lambda_0^2 [ {\rm Tr}(\alpha^2)-2
{\rm Tr}(\alpha)], \nonumber \\
{\rm Tr}[(\lambda_{x,v}-\lambda_0)^3]&=&2 \lambda_0^3 [ 3 {\rm Tr}(\alpha^2)-
{\rm Tr}(\alpha^3)],\nonumber \\
{\rm Tr}[(\lambda_{x,v}-\lambda_0)^4]&=&2 \lambda_0^4 [ {\rm Tr}(\alpha^4)- 4
{\rm Tr}(\alpha^3) +2 {\rm Tr}(\alpha^2)]. \label{4.19}
\end{eqnarray}
This is very convenient when
one truncates the momentum expansion as we will see in a moment.
We will denote by $\triangle_v^i U_x$ the coefficients in the expansion of
$\triangle_v U_x$ in powers of $a$. Relations
(\ref{4.16}) and (\ref{61}) imply that the
lowest order term in this expansion is the linear term:
\begin{equation}
\triangle_v U_x \equiv a ( \triangle_v^1 U_x) + \frac{a^2}{2} ( \triangle_v^2
U_x) + \frac{a^3}{6} ( \triangle_v^3 U_x)+ \ldots \label{4.20}
\end{equation}
Since the expansion of $\alpha$ starts with terms which are
quadratic in $a$, we can neglect terms proportional to ${\rm Tr}(\alpha^3)$,
${\rm Tr}(\alpha^4)$ in equations (\ref{4.19}), and keep only those
proportional to ${\rm Tr}(\alpha)$ and
${\rm Tr}(\alpha^2)$.

Next we define the
right and left handed external sources analogously to the continuum case:
\begin{equation}
r_v\equiv v_v+a_v,\quad l_v\equiv v_v-a_v.
\end{equation}
Any
variable which under the lattice symmetry transforms in the same way as
the link $v_{ij}^{\alpha}$ on which it is defined will from now on be
denoted by a subscript $v$. For example:
\begin{equation}
a_v \equiv \frac{a_i + \alpha a_j}{\sqrt{2}},\quad a_v^2\equiv (a_v)^2,
\;{\rm etc.}\; \label{52}
\end{equation}

Partial derivatives in the direction of the link $v_{ij}^{\alpha}$ are
defined similarly:
\begin{equation}
\partial_v a \equiv \frac{\partial_i a + \alpha \partial_j a}{\sqrt{2}},
\quad \partial_v^2 a \equiv \partial_v (\partial_v a),
\;{\rm etc.}\;
\end{equation}
These derivatives appear in the Taylor series expansion of
$r_{x+v,v}$ and $R_{x+v}$ in powers of a:
\begin{eqnarray}
r_{x+v,v}&=&r_v + a (\partial_v r_v) + \frac{a^2}{2}(\partial_v^2 r_v) +
\ldots ,\nonumber \\
R_{x+v}&=&R + a (\partial_v R) + \frac{a^2}{2}(\partial_v^2 R) +
\ldots,
\end{eqnarray}
and similarly for $l_{x+v,v}$ and $L_{x+v}$\footnote{Here we are using
the convention that, when there is no subscript $x$, a variable (or its
derivative), is evaluated at $x$.}.
By substituting the previous Taylor series expansions into equations
(\ref{4.16}) and (\ref{4.17}) one can derive the laws of
transformation of $r_v$ and $l_v$:
\begin{eqnarray}
r_v & \rightarrow& R r_v R^{\dagger}+ i R (\partial_v R^{\dagger}),
\nonumber \\
l_v & \rightarrow& L l_v L^{\dagger}\,\:+ i L (\partial_v L^{\dagger}).
\label{4.24}
\end{eqnarray}
Since this is true for any $v_{ij}^{\alpha}$ (\ref{17}), in the classical
continuum limit (\ref{3.9}) this law of transformation becomes the usual
law of transformation of gauge fields (\ref{2.6}).
Let us now define $D_v U$ on the link $v$ in analogy with the continuum
covariant derivative:
\begin{equation}
D_v U \equiv \partial_v U - i r_v U + i U l_v.
\end{equation}

In the classical continuum limit, $D_v U$ reduces to its continuum
counterpart (\ref{2.11}). Moreover,
from the transformation laws (\ref{4.24}), one can see that $D_v U$
transforms covariantly on the lattice:
\begin{equation}
D_v U \rightarrow R (D_v U) L^{\dagger}.
\end{equation}
We now expand $\triangle_v U_x$ in a Taylor series in $a$ by a
straightforward but tedious calculation and then re-express coefficients
in this expansion, defined in (\ref{4.20}), in terms of the lattice
covariant derivatives $D_v U$:\footnote{Here by $D_v^2 U$ we mean
$D_v(D_v U)$ etc.}
\begin{eqnarray}
\triangle_v^1 U_x &=& - D_v U, \nonumber \\
\triangle_v^2 U_x &=& - D_v^2 U - 2 i r_v D_v U,\nonumber \\
\triangle_v^3 U_x &=& - D_v^3 U - 3 i r_v D_v^2 U - 3 i (\partial_v r_v)
D_v U + 3 r_v^2 D_v U.
\end{eqnarray}
Although not all coefficients in the expansion of $\triangle_v U$ are
covariant under chiral transformations, ${\rm Tr}(\alpha)$
and ${\rm Tr}(\alpha^2)$ are manifestly covariant!
\begin{equation}
{\rm Tr}(\alpha)= {\rm Tr}\{a^2 (D_v U)^{\dagger} (D_v U) + a^4 [
\frac{1}{6} (D_v U)^{\dagger} (D_v^3 U) + \;{\rm h.c.}\; +
\frac{1}{4} (D_v^2 U)^{\dagger} (D_v^2 U) ] \},
\end{equation}
\begin{equation}
{\rm Tr}(\alpha^2)={\rm Tr}[ a^4 (D_v U)^{\dagger} (D_v U)
(D_v U)^{\dagger} (D_v U) ].
\end{equation}

In appendix B, we derive results for the summation of
tensors of rank two and four over the twelve links in the positive
direction:
\begin{equation}
\sum_v a_v b_v = 3 \sum_{\mu} a_{\mu} b^{\mu},\label{610}
\end{equation}
\begin{equation}
\sum_v a_v b_v c_v d_v = \frac{1}{2}\sum_{\mu,\nu}(a_{\mu} b^{\mu}
c_{\nu} d^{\nu} +a_{\mu} c_{\nu} b^{\mu} d^{\nu} +a_{\mu} d_{\nu}
b^{\nu} c^{\mu} ).\label{620}
\end{equation}
By using these expressions, summing over links and Wick rotating back to
Minkowski space, we get manifestly
Lorentz and gauge invariant expressions for the traces:
\begin{eqnarray}
\sum_v {\rm Tr}(\alpha) &=& {\rm Tr}\{ \frac{1}{12}
[(D_{\mu} U)^{\dagger} (D^{\mu}
D_{\nu} D^{\nu} U) + (D_{\mu} U)^{\dagger} (D^{\nu}
D^{\mu} D_{\nu} U) + (D_{\mu} U)^{\dagger} (D^{\nu}
D_{\nu} D^{\mu} U) \nonumber \\ +\;{\rm h.c}\;]
&+& \frac{1}{8} [(D^{\mu}D_{\mu} U)^{\dagger}(D^{\nu}D_{\nu} U) +
(D_{\mu}D_{\nu} U)^{\dagger}(D^{\mu}D^{\nu} U) +
(D_{\mu}D_{\nu} U)^{\dagger}(D^{\nu}D^{\mu} U) ]\},\nonumber \\
\sum_v {\rm Tr}(\alpha^2) &=& \frac{1}{2} {\rm Tr} [
(D_{\mu} U)^{\dagger} (D^{\mu} U) (D_{\nu} U)^{\dagger} (D^{\nu} U)
+
(D_{\mu} U)^{\dagger} (D_{\nu} U) (D^{\mu} U)^{\dagger} (D^{\nu} U)
\nonumber \\ & & +
(D_{\mu} U)^{\dagger} (D_{\nu} U) (D^{\nu} U)^{\dagger} (D^{\mu} U)].
\end{eqnarray}
The last step is to substitute these expressions back into (\ref{4.20}).
The action
on the lattice gives the Lagrangian after dividing with the volume
of space-time on the BCH lattice:
\begin{equation}
a^4 \sum_x = 2 \int d^4 x
\end{equation}
We can transform all the terms of order $p^4$ into the canonical basis of
\cite{gl2} using the property of SU(3) matrices derived there:
\begin{eqnarray}
{\rm Tr}[(D_{\mu} U)^{\dagger}(D_{\nu} U)(D^{\mu} U)^{\dagger}(D^{\nu} U)]&=&
-2 {\rm Tr}[(D^{\mu} U)^{\dagger}(D_{\mu} U)(D^{\nu} U)^{\dagger}(D_{\nu} U)]
\nonumber \\
+\frac{1}{2} \{ {\rm Tr}[(D^{\mu} U)^{\dagger} (D_{\mu} U)]\}^2 &+&
{\rm Tr}[(D^{\mu} U)^{\dagger}(D^{\nu} U)]
{\rm Tr}[(D_{\mu} U)^{\dagger}(D_{\nu} U)],
\end{eqnarray}
as well as the equation of motion at order $p^2$:
\begin{equation}
(D^{\mu}D_{\mu} U) U^{\dagger}  - U (D^{\mu}D_{\mu} U)^{\dagger}
=\chi U^{\dagger} - U \chi^{\dagger} - \frac{1}{3} {\rm Tr}(\chi
U^{\dagger} - U \chi^{\dagger} ).
\end{equation}

We have collected these results in Table 1. and compared them with
the coefficients which have been determined experimentally at the scale
of $m_{\eta}$ in \cite{gl2}
\begin{table}
\begin{center}
\begin{tabular}{|c|rc|rcl|}\hline
       &{\bf theory}&&\multicolumn{3}{c|}{{\bf experiment}} \\ \hline
$L_{1}$&0.5 $(F_{0}a)^{2}$&&0.9 &$\pm$& 0.3\\
$L_{2}$&1.1 $(F_{0}a)^{2}$&&1.7 &$\pm$& 0.7\\
$L_{3}$&10.4 $(F_{0}a)^{2}$&&-4.4 &$\pm$& 2.5\\
$L_{4}$&0 $(F_{0}a)^{2}$&&0 &$\pm$& 0.5\\
$L_{5}$&0 $(F_{0}a)^{2}$&&2.2 &$\pm$& 0.5\\
$L_{6}$&0 $(F_{0}a)^{2}$&&0 &$\pm$& 0.3\\
$L_{7}$&0.9 $(F_{0}a)^{2}$&&-0.4 &$\pm$& 0.15\\
$L_{8}$&2.6 $(F_{0}a)^{2}$&&1.1 &$\pm$& 0.3\\
$L_{9}$&-6.7 $(F_{0}a)^{2}$&&7.4 &$\pm$& 0.2\\
$L_{10}$&-6.7 $(F_{0}a)^{2}$&&-5.7 &$\pm$& 0.3\\ \hline
\end{tabular}
\end{center}
\caption{Comparison of the strong coupling expansion with experimental
results from \protect{\cite{gl2}},\protect{\cite{meissner}}.
The left column contains our results from the strong
coupling expansion. The right column contains the experimental values
\protect{ $L_i^r(m_{\eta})$}. All numbers are in units of
\protect{$10^{-3}$}. }
\end{table}
\section{Beyond the strong coupling regime}

\hspace{6mm}We have chosen to express the entries in Table 1. in terms of a
common factor $(F_0a)^2$ because of a point we will be making below, but
the value of this quantity is also determined by the strong coupling
expansion, from the $O(p^2)$ terms in the lattice chiral Lagrangian, and
is given by $2.1 N$.  With 3 colors the entries in the table come out
about one order of magnitude higher than the values derived from
experiment.  We take this to be an indication that the strong
coupling limit produces too tight a binding between quark and
antiquark, which in turn leads to an unacceptably high value when $F_0$
is expressed in terms of $a^{-1}$ (or, equivalently, to too large a
value for $a$).  If we allowed ourselves to set $a^{-1} \approx F_0$,
which would be a more reasonable scale for a strong coupling
calculation, the theoretical predictions for several
coefficients would compare rather well with the experimental
results.  Of course we should not make too much of this agreement,
because we expect the values of the coefficients to depend on
detailed dynamical features of the interactions which may not
be well reproduced by the strong coupling approximation.  What is much
more important is the demonstration, through the strong coupling
expansion, of the main point we wanted to make, namely that one can
formulate an effective chiral Lagrangian on the lattice and that
this can be the vehicle for the derivation of the parameters of the
continuum chiral Lagrangian.

Another point that should be mentioned is that the expansion for
small lattice momenta that we have used to derive the coefficients
corresponds to using the lattice chiral Lagrangian in the tree
approximation.  We have calculated a few of the corrections that would
be induced by one loop diagrams and have found these to be small.
However, the same factor of $(F_0 a)^2$ which we discussed above enters
also as a coefficient in the denominator of the loop diagrams, so that the
statement that the loop corrections are small should be taken with
caution. They would become larger if one could assign to $F_0 a$ the
smaller value which we advocated above. Nevertheless, even if the loop
corrections turned out to be large in a more realistic calculation,
this would only represent a technical problem and not a conceptual
difficulty for the whole approach.  The very important fact is that,
because we are working with a regularized theory, all physical
quantities are well defined and finite.  Moreover, since the lattice system
also exhibits spontaneous breaking of chiral symmetry and
Goldstone bosons in the limit of vanishing quark mass, it will
always be possible to perform an expansion for small lattice
momenta.  The coefficients of such expansion are in any case
well defined quantities.  Thus the question is whether they can be
calculated by a perturbative expansion of the chiral effective theory,
which would be more convenient, or whether, failing such possibility,
one will have to resort to numerical techniques. Nevertheless,
even in the latter case, one would still be dealing with a bosonic
system, and would therefore avoid the need of simulating the quantum
fluctuations of fermionic variables, which represents today the major
difficulty one faces in the implementation of QCD numerical simulations.

Finally, in order to obtain reliable values for the coefficients one
should perform the calculation in the range of values of the lattice
coupling constant ($\beta\approx 5.7-6$, including quark
degrees of freedom in the simulation) where one witnesses the onset of
scaling to the continuum.  Such a calculation can at present only be
done by numerical techniques.  The way we envisage it could be carried
out would be by assuming a sufficiently large set of couplings for
the lattice effective theory and fixing them through the matching of
an overcomplete set of expectation values.  This is very similar to
procedures commonly used in numerical studies of renormalization
group transformations.  The crucial point is that, because the
lattice effective theory already accounts for the long range excitations
of QCD, the matching should only require a reasonably small lattice
size, of the order of the inverse $\rho$ mass.  The two theories
(the one derived from the original QCD Lagrangian and the effective
theory) should produce exactly the same values for the observables on
any lattice size, because they are mathematically equivalent.
Working with a small lattice, we are confident, one would be able to fix
the parameters of the lattice effective theory with a good degree
of accuracy and would thus establish a solid base for the
subsequent determination of the parameters of the continuum
effective chiral Lagrangian.

{\bf Acknowledgments}
We would like to thank Richard Brower, Christophe Bruno,
Sekhar Chivukula, Andrew Cohen, Gerhard Ecker, Vic Emery, Barry Holstein,
Dimitris Kominis, Yue Shen, Jan Smit, Akira Ukawa and Steven Weinberg for
useful and stimulating discussions.
\vspace{1cm}
\appendix{{\large\bf Appendix A. Goldstone bosons}}
\vspace{6mm}

In this appendix, we will show that the fields $\pi_x$ are
Goldstone bosons, corresponding to the spontaneous breaking of
the chiral symmetry.

It was shown in \cite{klus}, \cite{kaws}, for the case of no external
sources on the hypercubical lattice, that the action in (\ref{3.29})
exhibits spontaneous chiral symmetry breaking:
\begin{equation}
U(n)\times U(n) \rightarrow U(n).
\end{equation}
In that context, $n$ represented the number of color triplets of staggered
fermions. Our situation is somewhat different: we start from
full-fledged fermions with Dirac, flavor and color indices and $n$
represents the number of flavors. Moreover, we have a different lattice
structure.

We obtain the inverse propagator in momentum\footnote{In this appendix,
for simplicity we take $a=1$.}
 space for $n^2$ fields
$\pi_x$ from expressions (\ref{4.11}), (\ref{41}),
(\ref{61}) and (\ref{4.19}) by expanding to the quadratic order in
powers of fields $\pi_x$:
\begin{equation}
D(p)=B[12-\sum_v \cos{(p \cdot v)} ] ,\label{A1}
\end{equation}
\begin{equation}
B=\frac{4 N}{F_0^2}[\lambda_0 F'(\lambda_0)+\lambda_0^2 F''(\lambda_0)].
\end{equation}
We can rewrite the momentum dependent part of $D(p)$ in terms of
momentum components $p_{\mu}$:
\begin{eqnarray}
\sum_v \cos{ (p \cdot v)} = &2& [\cos{\frac{p_4}{\sqrt{2}}}
(\cos{\frac{p_1}{\sqrt{2}}} + \cos{\frac{p_2}{\sqrt{2}}}
+\cos{\frac{p_3}{\sqrt{2}}} )+\nonumber \\
 & & \cos{\frac{p_3}{\sqrt{2}}}
(\cos{\frac{p_1}{\sqrt{2}}} +\cos{\frac{p_2}{\sqrt{2}}} )
+\cos{\frac{p_1}{\sqrt{2}}} \cos{\frac{p_2}{\sqrt{2}}}].
\end{eqnarray}
By substituting this back into the expression (\ref{A1})
one can immediately see that the propagator $D^{-1}(p)$ has a massless
pole $p_4=i m =0$ at the origin ${\bf p}=0$ for each of the $n^2$ fields
contained in the matrix $\pi_x$.
\vspace{1cm}
\appendix{{\large\bf Appendix B. Lorentz covariance}}
\vspace{6mm}

In this appendix we derive relation (\ref{610}). From
the definition (\ref{52}):
\begin{eqnarray}
\sum_v a_v b_v &=& \sum_{i,j,\alpha} \frac{a_i + \alpha a_j}{\sqrt{2}}
\; \frac{b_i + \alpha b_j}{\sqrt{2}}\nonumber\\
&=&\frac{1}{2}\sum_{i,j,\alpha}[a_i b_i+a_j b_j +\alpha (a_i b_j+a_j
b_i)].
\end{eqnarray}
When summing over $\alpha=\pm1$, terms proportional to $\alpha$ cancel
and the others get a factor of $2$:
\begin{equation}
\sum_v a_v b_v = \sum_{i,j}(a_i b_i+a_j b_j).
\end{equation}
Finally, we perform the sum over $1\leq i < j \leq 4$ and obtain:
\begin{equation}
\sum_v a_v b_v = 3\sum_{\mu}a_{\mu} b^{\mu}.
\end{equation}
Relation (\ref{620}) can be proven in the same way.

\end{document}